\documentclass[aps,prl,twocolumn,superscriptaddress,showpacs,showkeys]{revtex4-2}

\usepackage{color}
\usepackage{graphicx}
\usepackage{dcolumn}
\usepackage{bm}
\usepackage{float}
\usepackage[mathlines]{lineno}
\usepackage{tabularx}
\usepackage[colorlinks,linkcolor=blue,anchorcolor=blue,citecolor=blue,urlcolor=blue,hyperindex,CJKbookmarks]{hyperref}
\usepackage{footnote}
\usepackage{amsmath}
\usepackage{txfonts}
\usepackage{mathptmx}
\usepackage{ragged2e}
\usepackage{booktabs,makecell, multirow, tabularx}
\usepackage{multirow}
\usepackage{braket}
\usepackage{gensymb}
\usepackage{array}
\usepackage{physics} 
\usepackage{textcomp}
\usepackage[T1]{fontenc}

\begin{document}

\title{Ultrasmall (3,0) carbon nanotube: a superconducting unit with strong electron-phonon coupling}

\author{Zhenfeng Ouyang}\affiliation{School of Physics and Beijing Key Laboratory of Opto-electronic Functional Materials $\&$ Micro-nano Devices, Renmin University of China, Beijing 100872, China}\affiliation{Key Laboratory of Quantum State Construction and Manipulation (Ministry of Education), Renmin University of China, Beijing 100872, China}

\author{Jing Jiang}\affiliation{School of Physics and Beijing Key Laboratory of Opto-electronic Functional Materials $\&$ Micro-nano Devices, Renmin University of China, Beijing 100872, China}\affiliation{Key Laboratory of Quantum State Construction and Manipulation (Ministry of Education), Renmin University of China, Beijing 100872, China}

\author{Jian-Feng Zhang}\affiliation{Center for High Pressure Science $\&$ Technology Advanced Research, Beijing 100193, China}

\author{Miao Gao}\affiliation{Department of Physics, School of Physical Science and Technology, Ningbo University, Zhejiang 315211, China}

\author{Kai Liu}\email{kliu@ruc.edu.cn}\affiliation{School of Physics and Beijing Key Laboratory of Opto-electronic Functional Materials $\&$ Micro-nano Devices, Renmin University of China, Beijing 100872, China}\affiliation{Key Laboratory of Quantum State Construction and Manipulation (Ministry of Education), Renmin University of China, Beijing 100872, China}

\author{Zhong-Yi Lu}\email{zlu@ruc.edu.cn}\affiliation{School of Physics and Beijing Key Laboratory of Opto-electronic Functional Materials $\&$ Micro-nano Devices, Renmin University of China, Beijing 100872, China}\affiliation{Key Laboratory of Quantum State Construction and Manipulation (Ministry of Education), Renmin University of China, Beijing 100872, China}\affiliation{Hefei National Laboratory, Hefei 230088, China}

\date{\today}

\begin{abstract}
Based on density-functional first-principles calculations, we study the electronic structures, lattice dynamics, and electron-phonon coupling (EPC) properties of the ultrasmall one-dimensional (3,0) carbon nanotube (CNT). Our studies reveal that the (3,0) CNT hosts characteristic collective phonon modes, in which the corresponding atomic vibrations can significantly modulate the electronic self-consistent potential and induce strong EPC. Our results further show that the (3,0) CNT exhibits robust phonon-mediated superconductivity under ambient pressure, with a superconducting transition temperature reaching $\sim$ 33 K ($\mu^*$ = 0.10). We also find that although hole doping enhances the density of states (DOS) near the Fermi level, it simultaneously hardens the strongly coupled phonon modes and suppresses the total EPC. This result indicates that phonons play a more crucial role in the superconductivity of ultrasmall CNTs rather than the DOS alone. Therefore, we suggest that the ultrasmall (3,0) CNT can serve as a superconducting unit hosting phonon modes with strong EPC, and propose a strategy for discovering new phonon-mediated superconductors by identifying superconducting units.

\end{abstract}


\maketitle

\textit{Introduction}. Since the discovery of superconductivity, searching for conventional superconductors with high superconducting transition temperatures ($T_c$) has remained a central topic in condensed matter physics. In conventional superconductors, metallization of $\sigma$-bonding electrons has long been regarded as one of the most effective routes toward high $T_c$~\cite{2015-44-07-001}, which successfully explains the superconductivity of MgB$_2$~\cite{CHOI200366}. Nevertheless, our recent study~\cite{PhysRevB.111.L140501} suggests that characteristic phonon modes with strong electron-phonon coupling (EPC) may play a more fundamental role in determining supercon-ducting pairing strength. In this context, identifying local structural motifs that can generate phonon modes with exceptionally strong EPC provides a possible strategy for discovering high-$T_c$ conventional superconductors, giving a complementary perspective to the conventional paradigm based on $\sigma$-bond metallization.

Motivated by this idea, we proposed a concept of ``superconducting unit'', referring to a local structural motif that hosts characteristic phonon modes with exceptionally large EPC~\cite{PhysRevB.111.L140501}. Representative examples include some characteristic phonon modes in several EPC superconductors, such as MgB$_2$~\cite{PhysRevB.65.132518}, H$_3$S~\cite{PhysRevB.108.094519}, and Li$_2$AuH$_6$~\cite{PhysRevB.111.L140501,rltl-vgzj}, where strong phonon vibrations induce significant modulations of the electronic self-consistent potential, leading to strong EPC. From this perspective, searching for superconducting units possessing characteristic phonon modes with strong EPC provides a new and promising strategy for exploring high-$T_c$ conventional superconductivity.

Low-dimensional carbon nanotubes (CNTs) provide an attractive platform for this purpose. Due to the light atomic mass of carbon and strong covalent bonding, carbon materials usually possess high phonon frequencies that may be favorable for EPC superconductivity. In particular, ultrasmall CNTs combine several advantageous features simultaneously, including dense bonding geometries, strong curvature-induced hybridization, and one-dimensional electronic structures with van Hove singularities near the Fermi level. These characteristics further make ultrasmall CNTs promising candidates for hosting phonon modes with strong EPC, as well as enhanced superconducting pairing strength.

Earlier experimental observations revealed that some large-diameter CNT systems exhibit superconductivity with $T_c$ below 20 K, including the 4-{\AA} single-walled CNTs embedded in a zeolite matrix ($T_c$ $\sim$ 15 K)~\cite{Tang2001} and the entirely end-bonded multiwalled CNTs ($T_c$ $\sim$ 12 K) ~\cite{PhysRevLett.96.057001}, while superconductivity of smaller-sized CNTs remains to be further investigated.

Theoretically, some previous studies investigated the electronic structures and EPC properties of various CNTs~\cite{PhysRevLett.72.1878,PhysRevLett.94.015503,PhysRevB.71.035429,PhysRevB.73.035413,LIANG20082288,Zhang2017,Ferrier_2010,WONG2017509,Odom1998,PhysRevB.65.155411}, while distinct conclusions were obtained depending on the tube diameter, chirality, and computational approaches~\cite{PhysRevLett.94.015503,PhysRevB.71.035429,PhysRevB.73.035413}. More recently, high-throughput calculations on element-doped CNTs predicted several superconducting candidates with relatively high superconducting $T_c$’s~\cite{doi:10.1021/acs.jpclett.1c02000}. Among them, La-doped (3,3) CNT shows the highest superconducting $T_c$ $\sim$ 29 K~\cite{doi:10.1021/acs.jpclett.1c02000}. However, most existing investigations mainly focus on CNTs with relatively large diameters, whereas studies regarding the EPC superconducting properties of ultrasmall CNTs are quite limited.

Within the ultrasmall CNTs, the one-dimensional (3,0) CNT is particularly intriguing because it lies near the stability limit of freestanding carbon nanotubes and possesses strong curvature effects~\cite{PhysRevLett.87.125502}. Such a special dense structural geometry is expected to significantly enhance phonon-induced modulations of the electronic structures, which may give rise to characteristic phonon modes with strong EPC. In this work, we systematically investigate the electronic structures, lattice dynamics, and EPC properties of one-dimensional (3,0) CNT. We identify several characteristic phonon modes with exceptionally strong EPC, among which a breathing mode exhibits a remarkably large mode-resolved EPC constant of $\lambda_{\bf{q}\nu}$ $\sim$ 6.64. Hole doping is found to increase the density of states (DOS) near the Fermi level by shifting the van Hove singularity. However, it simultaneously hardens these characteristic phonon modes and reduces the total EPC constant, demonstrating that superconducting pairing in this system is primarily governed by the characteristic phonon modes than the DOS alone. By solving the anisotropic Migdal-Eliashberg equations~\cite{PhysRevB.87.024505,RevModPhys.89.015003}, the superconducting $T_c$ of pristine (3,0) CNT is $\sim$ 33 K ($\mu^*$ = 0.10). These results indicate that the ultrasmall (3,0) CNT is a promising phonon-mediated superconducting unit.

\begin{figure}[b]
\centering
\includegraphics[width=8.6cm]{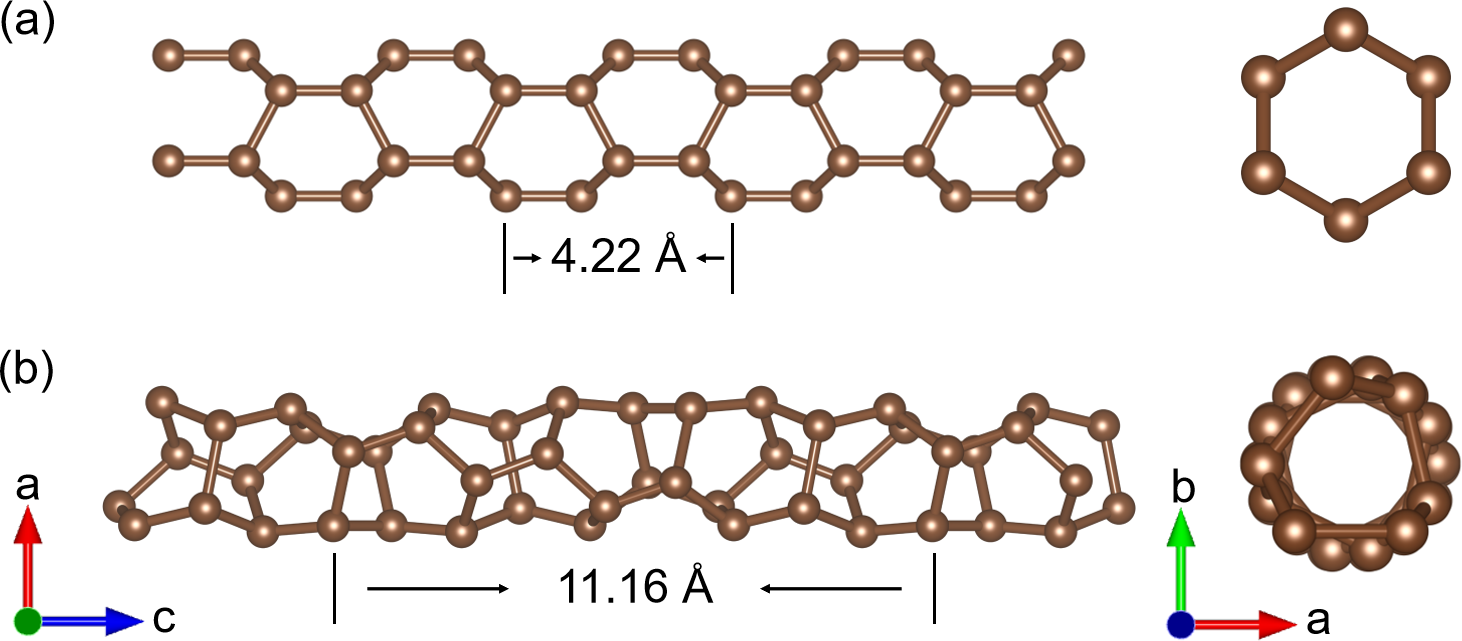}
\caption{Crystal structures for one-dimensional (a) (3,0) and (b) (2,1) carbon nanotubes, respectively.}
\label{fig:Fig1}
\end{figure}

\textit{Methods}. The density functional theory (DFT) calculations were performed with the QUANTUM ESPRESSO~\cite{Giannozzi_2009} package. The generalized gradient approximation of the Perdew-Burke-Ernzerhof formula~\cite{PhysRevLett.77.3865} for the exchange and correlation functional and the optimized norm-conserving Vanderbilt pseudopotential~\cite{PhysRevB.88.085117} were used. The charge densities were determined self-consistently on an unshifted 1$\times$1$\times$40 $\bf{k}$-point grid with a Methfessel-Paxton smearing of 0.02 Ry~\cite{PhysRevB.40.3616}. The kinetic energy cutoff and the charge density cutoff were set to 80 and 320 Ry, respectively. 

Based on the density-functional perturbation theory (DFPT)~\cite{RevModPhys.73.515}, the dynamical matrices and the perturba-tion potential were calculated on a 1$\times$1$\times$6 $\bf{q}$-point grid. Maximally localized Wannier functions were constructed by using the Wannier90 code~\cite{Pizzi_2020}, and the converged EPC constant $\lambda$ was determined in the fine electron (1$\times$1$\times$360) and phonon (1$\times$1$\times$360) grids with the electron-phonon Wannier (EPW) package~\cite{PONCE2016116}. The Dirac $\delta$ functions for electrons and phonons were smeared out by the Gaussian function with widths of 30 and 0.5 meV, respectively. The sum over Matsubara frequencies was truncated with $\omega_{c}$ = 1.3 eV when solving the anisotropic Migdal-Eliashberg equations~\cite{PhysRevB.87.024505,RevModPhys.89.015003}.

\textit{Electronic structures of ultrasmall carbon nanotubes}. In Fig.~\ref{fig:Fig1}, we show the crystal structures of the one-dimensional ultrasmall (3,0) and (2,1) CNTs, which were proposed to lie close to the structural stability limit of freestanding CNTs while simultaneously exhibiting extremely strong curvature effects. More importantly, the low-energy electronic states of such dense crystal structures are sensitive to phonon vibrations, making ultrasmall CNTs promising candidates for hosting strong EPC.

According to the conventional tight-binding rule, a ($m$,$n$) CNT is expected to be metallic (semiconducting) when $m$-$n$ = 3$P$ (3$P$ $\pm$ 1), where $P$ is an integer~\cite{PhysRevLett.68.1579,RevModPhys.79.677}. Therefore, the (3,0) CNT is expected to exhibit metallic behavior. As shown in Fig.~\ref{fig:Fig1}(a), the C-C bonds of the (3,0) CNT form an armchair arrangement along the $c$ axis and a zigzag arrangement along the $a$ axis. As a comparison, the ultrasmall (2,1) CNT, which possesses a chiral structure without mirror symmetry, forms a spiral arrangement along the tube axis, as shown in Fig.~\ref{fig:Fig1}(b). According to the conventional rule~\cite{RevModPhys.79.677}, the (2,1) CNT is expected to be semiconducting.

\begin{figure}[h]
\centering
\includegraphics[width=8.6cm]{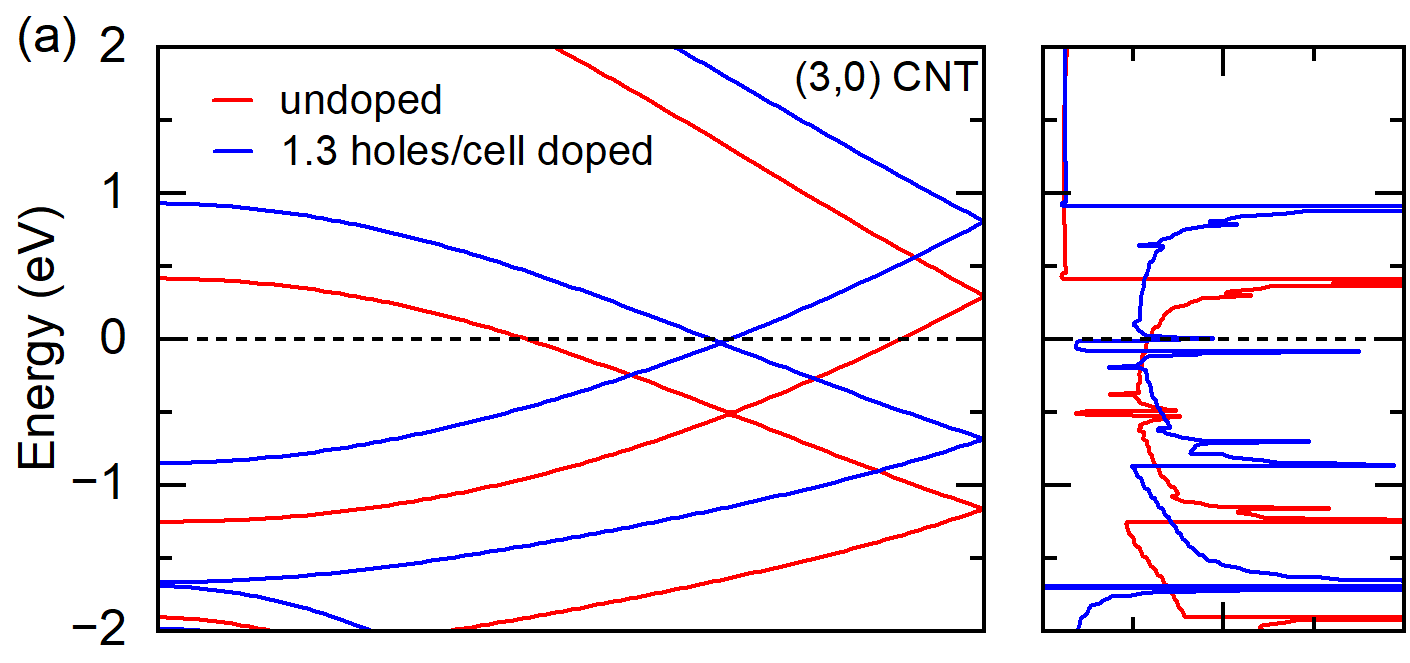}
\includegraphics[width=8.6cm]{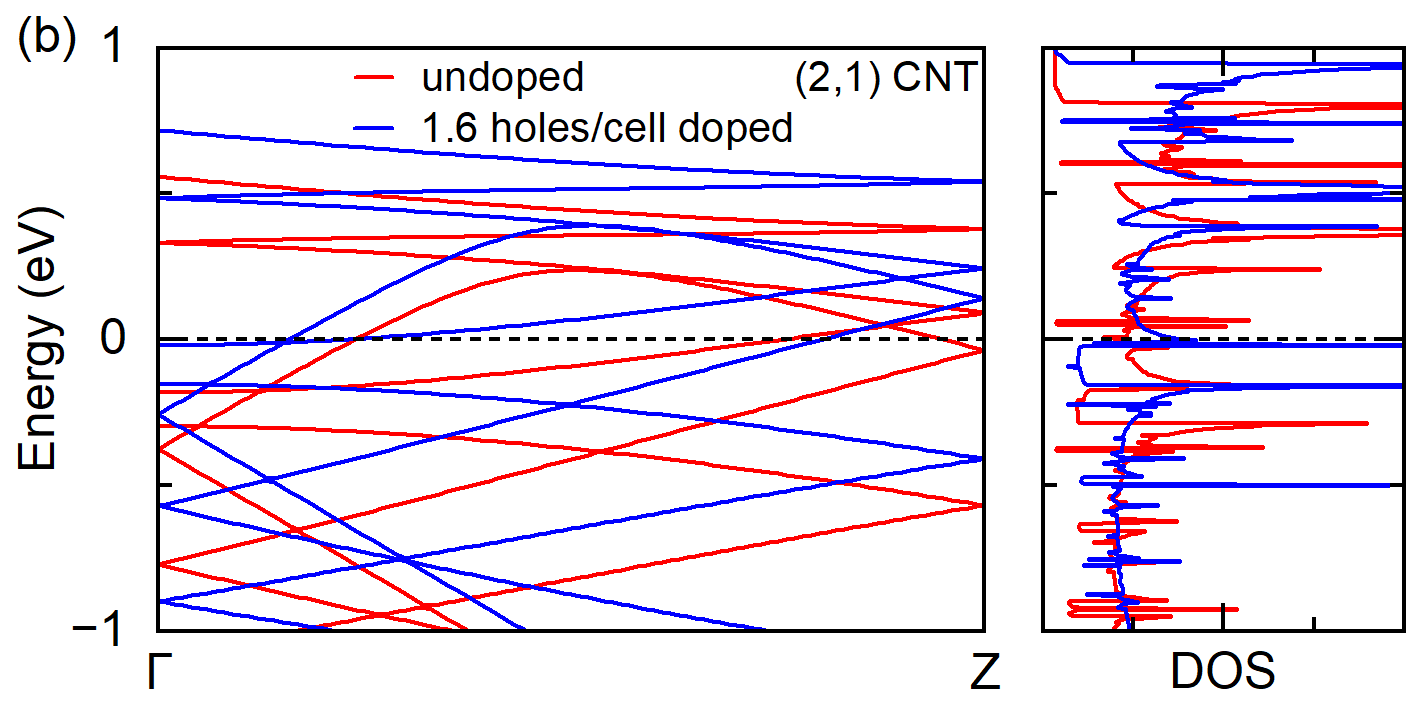}
\caption{Electronic structures for (a) undoped and 1.3 holes/cell doped (3,0) and (b) undoped and 1.6 holes/cell doped (2,1) carbon nanotubes, respectively. The Fermi level is set to zero.}
\label{fig:Fig2}
\end{figure}

However, as shown in Fig.~\ref{fig:Fig2}, our DFT calculations reveal that both the (3,0) and (2,1) CNTs exhibit metallic electronic structures. In particular, the metallic behavior of the (2,1) CNT indicates that the conventional classification of CNTs can break down in the ul-trasmall-diameter case, where strong curvature-induced $\sigma^*$-$\pi^*$ hybridization~\cite{PhysRevLett.72.1878} can reconstruct the low-energy electronic states to be metallic. This is conducive to the emergence of superconductivity in ultrasmall CNTs. In addition, pronounced peaks are found in the DOS near the low-energy region in both (3,0) and (2,1) CNTs, which are attributed to the van Hove singularities associated with the extreme points of bands of low-dimensional systems. Such low-dimensional electronic structures with strong $\sigma^*$-$\pi^*$ hybridization may be sensitive to some special phonon vibrations, which is favorable for strong EPC.
Moreover, the low-energy electronic structures can be further modulated by chemical doping. Here, we also simulate hole doping within the rigid-band approximation by modifying the total electron number while keeping the overall band topology nearly unchanged to investigate the evolution of the low-energy electronic structures of ultrasmall CNTs. This situation also corresponds to experiments, where hole doping in carbon-based materials is usually realized by substituting C with B atoms. Since elements B and C possess very similar chemical properties, such substitution can effectively introduce hole carriers while largely preserving the overall electronic structure. As shown in Fig.~\ref{fig:Fig2}, our calculated results show that after applying 1.3 and 1.6 holes/cell doping to the (3,0) and (2,1) CNTs, respectively, the van Hove singularities can be shifted toward the Fermi level, and the low-energy electronic states are significantly reconstructed. These results indicate that hole doping can effectively tune the low-energy electronic structures of ultrasmall CNTs, providing an additional route for investigating the EPC properties in these systems.

\begin{figure}[h]
\centering
\includegraphics[width=8.6cm]{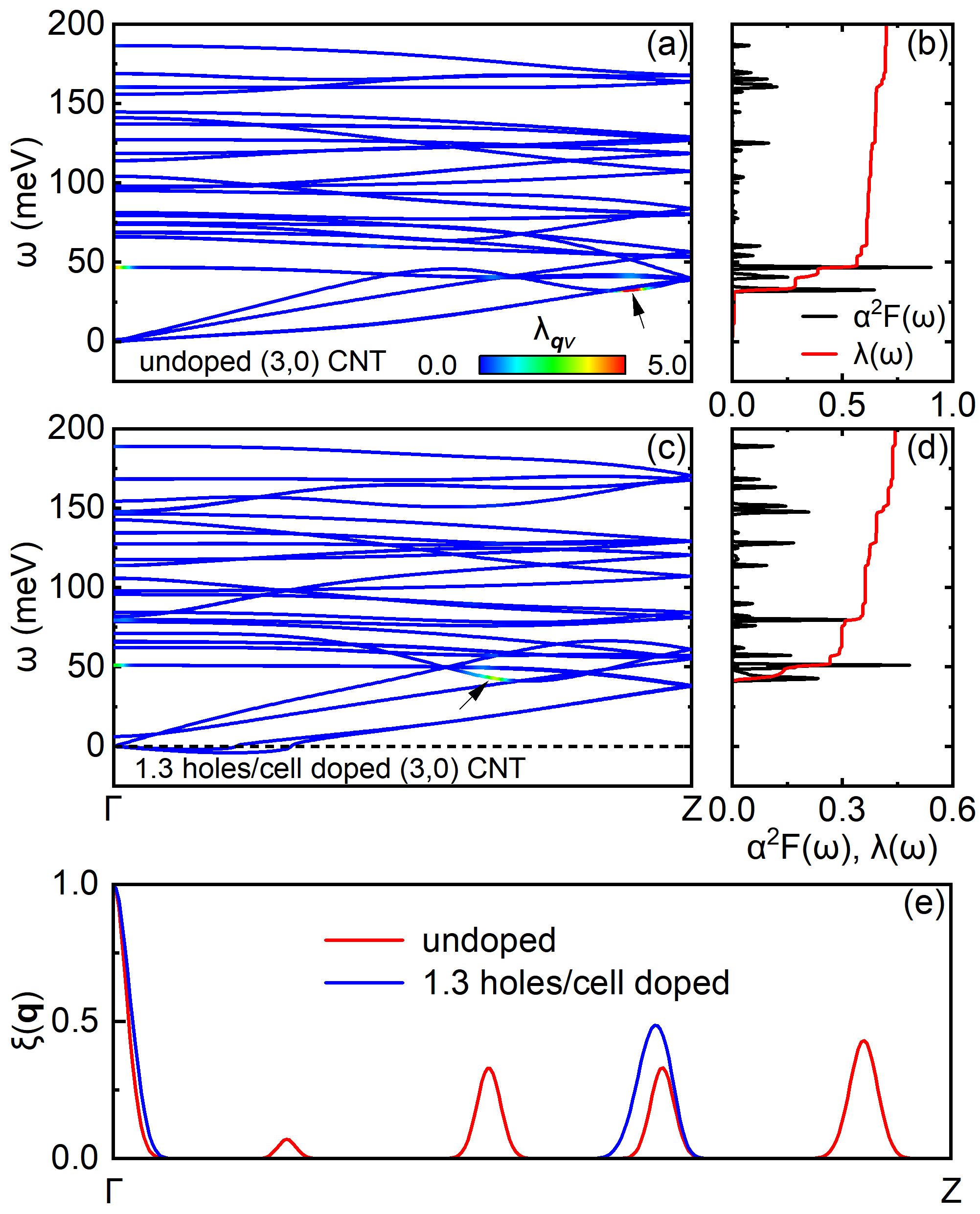}
\caption{Phonon spectrum with a color representation of $\lambda_{\bf{q}\nu}$, Eliashberg spectral function $\alpha^2F(\omega)$ and accumulated $\lambda(\omega)$ for (a-b) undoped and (c-d) 1.3 holes/cell doped (3,0) carbon nanotubes. The scale of $\alpha^2F(\omega)$ is omitted for clarity. (e) Fermi-surface nesting function $\xi(\bf{q})$ of undoped and 1.3 holes/cell doped (3,0) carbon nanotubes. $\xi(\bf{q})$ is normalized by the corresponding value of $\xi(\Gamma)$.}
\label{fig:Fig3}
\end{figure}

\textit{Electron-phonon coupling of ultrasmall (3,0) carbon nanotubes}. We next investigate the lattice dynamics and EPC properties of the ultrasmall (3,0) CNT. Starting from the undoped case, the calculated phonon spectrum of the (3,0) CNT is shown in Fig.~\ref{fig:Fig3}(a). No imaginary phonon mode is observed in the whole Brillouin zone, indicating that the undoped (3,0) CNT is dynamically stable under ambient pressure. Further EPC calculations identify several optical phonon modes with exceptionally strong EPC in the frequency range of 30-50 meV, where the corresponding mode-resolved EPC constants $\lambda_{\bf{q}\nu}$ exceed 2. These phonon modes give rise to pronounced peaks in the Eliashberg spectral function $\alpha^2F$($\omega$), as shown in Fig.~\ref{fig:Fig3}(b). The total EPC constant $\lambda$ of the undoped (3,0) CNT is calculated to be 0.70, among which nearly 80\% of the accumulated EPC strength originates from the 30-50 meV frequency region. These results indicate that a few characteristic phonon modes strongly dominate the EPC of the ultrasmall (3,0) CNT.

We also study the lattice dynamics and EPC properties of 1.3 holes/cell doped (3,0) CNT. As shown in Figs.~\ref{fig:Fig3}(c) and (d), small imaginary acoustic phonon modes emerge near the $\Gamma$ point after hole doping, which is a common feature in low-dimensional materials and may be further stabilized under substrate effects, as observed experimentally in 4-{\AA} single-walled CNTs embedded in zeolite matrices~\cite{Tang2001}. Interestingly, although hole doping shifts the van Hove singularities toward the Fermi level and enhances the low-energy density of states, the total EPC constant $\lambda$ is reduced from 0.70 to 0.44. Meanwhile, we also notice that those phonon modes with strong EPC exhibit hardening after hole doping, where the low-frequency phonon mode around 30 meV also shifts away from the Z point, as marked by the black arrows in Figs.~\ref{fig:Fig3}(a) and (c). These findings imply that EPC evolution in (3,0) CNT cannot be understood from the electronic aspect, such as DOS, alone. On the contrary, phonon would be a more crucial factor.

To further understand the momentum dependence of the phonon modes with strong EPC, we calculate the Fermi-surface nesting function $\xi$($\bf{q}$), as shown in Fig.~\ref{fig:Fig3}(e). Under both undoped and doped conditions, the strong EPC phonon modes emerge near the characteristic nesting vectors where $\xi$($\bf{q}$) exhibits pronounced peaks. According to the definitions of mode-resolved EPC constant $\lambda_{\bf{q}\nu}$ and nesting function $\xi$($\bf{q}$)~\cite{RevModPhys.89.015003}, which read:
\begin{align}
\lambda_{\bf{q}\nu} =\frac{2}{\hbar{N(0)}{N_{\bf{k}}}}\Sigma_{nm\bf{k}}\frac{1}{\omega_{\bf{q}\nu}}|g^{nm}_{\bf{k},\bf{q}\nu}|^2\delta(\epsilon^n_{\bf{k}})\delta(\epsilon^m_{\bf{k+q}})
\end{align}
and
\begin{align}
\xi{(\bf{q})} =\frac{1}{{N(0)}{N_{\bf{k}}}}\Sigma_{nm\bf{k}}\delta(\epsilon^n_{\bf{k}}-\epsilon_{{F}})\delta(\epsilon^m_{\bf{k+q}}-\epsilon_{{F}}).
\end{align}
$N$(0) is the electronic density of states at the Fermi level, $N_{\bf{k}}$ is the total number of $\bf{k}$ points, $\omega_{\bf{q}\nu}$ is the phonon frequency, and $g^{nm}_{\bf{k},\bf{q}\nu}$ is the EPC matrix element. The indices $n$, $m$, and $\nu$ denote the electronic bands and phonon modes, respectively. In one-dimensional systems, the Fermi surfaces reduce to discrete Fermi points. It can be naturally understood that the dominant EPC scattering processes can only exist at the nesting vectors connecting these Fermi points. For the 1.3 holes/cell doped (3,0) CNT, hole doping changes the topology of the Fermi surfaces, as well as the nesting vectors, which further gives rise to the stiffening of those phonon modes with strong EPC. As a consequence, the total EPC constant $\lambda$ is reduced from 0.70 to 0.44 after hole doping. These results highlight that the strong EPC of ultrasmall (3,0) CNT is primarily governed by the characteristic phonon modes.

\textit{Characteristic phonon modes with strong electron-phonon coupling}. To further clarify the microscopic origin of the strong EPC in the ultrasmall (3,0) CNT, we next analyze the characteristic phonon modes with exceptionally large mode-resolved EPC constants. According to the definition of the EPC matrix element~\cite{RevModPhys.89.015003}:
\begin{align}
g_{nm\nu}({\bf{k, q}}) = \langle{u}_{m, {\bf{k+q}}}|\Delta_{{\bf{q}}\nu}V^{{\bf{KS}}}|u_{n, {\bf{k}}}\rangle,
\end{align}
strong EPC requires substantial phonon-induced modulations of the electronic self-consistent potential $\Delta_{\bf{q}\nu}V^{\bf{KS}}$ experienced by electrons, where $u_{n, \bf{k}}$ and ${u}_{m, \bf{k+q}}$ denote the Bloch-periodic parts of the Kohn-Sham wave functions. Therefore, collective lattice distortions that can strongly modify the electronic self-consistent potential are expected to generate exceptionally large EPC.

\begin{figure}[h]
\centering
\includegraphics[width=8.6cm]{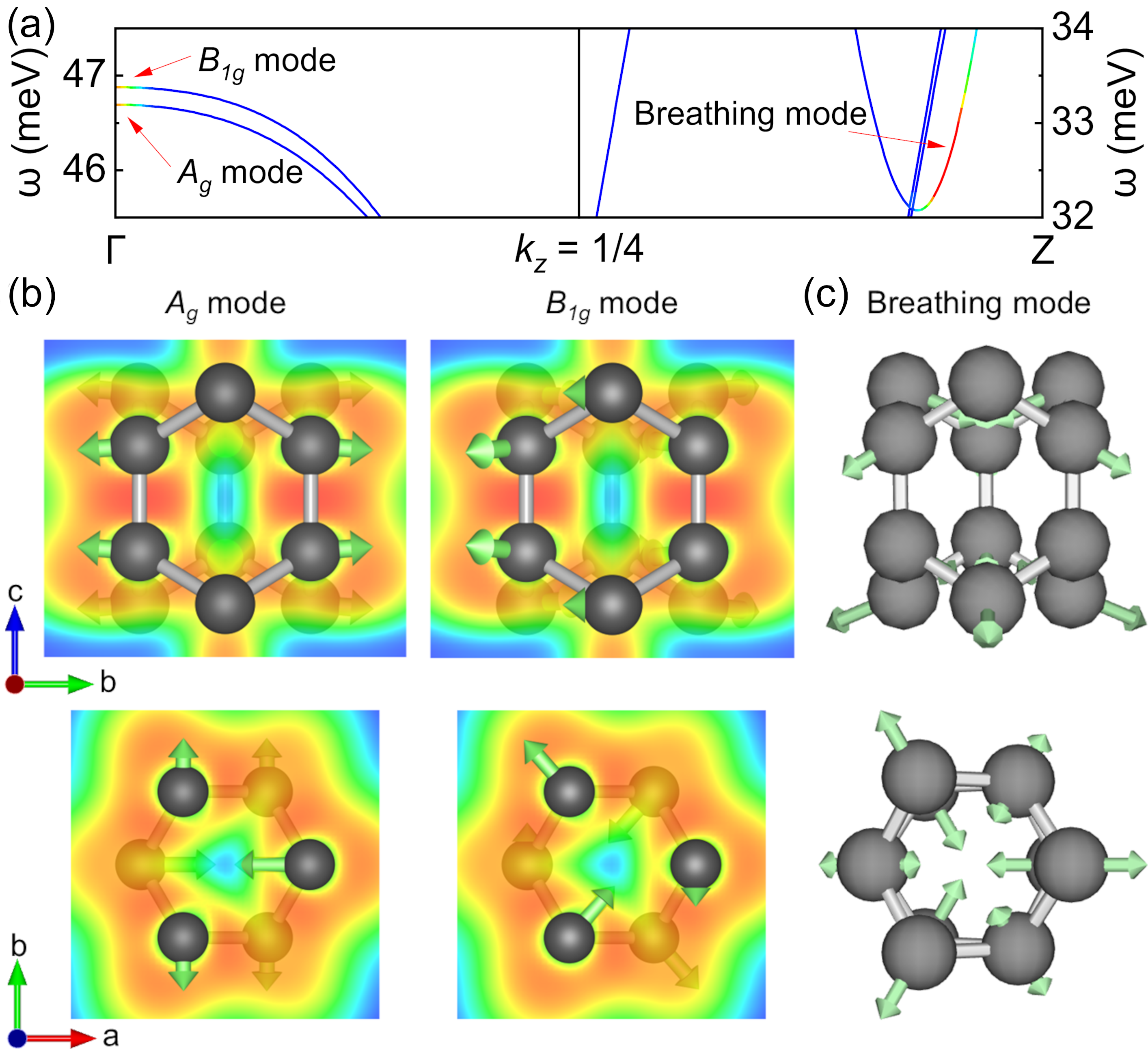}
\caption{(a) Zoom-in pictures of phonon spectra with $\lambda_{\bf{q}\nu}$ of undoped (3,0) carbon nanotube. Red arrows denote three different phonon modes with strong EPC. (b) Electron localization functions of undoped (3,0) carbon nanotube with vibration patterns of the $A_g$ and $B_{1g}$ phonon modes at the $\Gamma$ point. (c) The snapshots for the vibrations of the breathing mode that is near the $Z$ point of undoped (3,0) carbon nanotube.}
\label{fig:Fig4}
\end{figure}

Therefore, we study several optical phonon modes near the $\Gamma$ and $Z$ points that exhibit remarkably strong EPC [as shown in Fig.~\ref{fig:Fig4}(a)]. Among them, two nearly degenerate phonon modes with $A_g$ and $B_{1g}$ symmetries are identified around 47 meV near the $\Gamma$ point. The corresponding vibration patterns are illustrated in Fig.~\ref{fig:Fig4}(b). In these modes, carbon atoms exhibit in-plane vibrations, which periodically stretch and compress the zigzag carbon hexagons along different directions. Such collective lattice distortions strongly modulate the local C-C bonding environment and therefore generate exceptionally large EPC strengths with $\lambda_{\bf{q}\nu}$ $\sim$ 4.26 and 4.22 for the $A_g$ and $B_{1g}$ modes, respectively. After 1.3 holes/cell doping, both the $\lambda_{\bf{q}\nu}$ of the $A_g$ and $B_{1g}$ phonon modes are weakened to $\sim$ 2.52.

In addition to the $\Gamma$-point modes, an even stronger EPC phonon mode is identified around the $Z$ point with a remarkably large mode-resolved EPC constant of $\lambda_{\bf{q}\nu}$ $\sim$ 6.64 and 3.42 for the undoped and 1.3 holes/cell doped cases, respectively. As shown in Fig.~\ref{fig:Fig4}(c), this phonon mode corresponds to a collective breathing distortion of the zigzag carbon hexagons, where the three carbon atoms at the same height move collectively toward or away from the center of the hexagon. Although vibrations of different zigzag carbon hexagons exhibit phase differences along the tube axis, all carbon atoms vibrate with nearly identical amplitudes. Such a characteristic breathing vibration induces substantial modulations of the electronic environment throughout the carbon network, giving rise to exceptionally strong EPC.

We also show the electron localization functions of the ultrasmall (3,0) CNT in Figs.~\ref{fig:Fig4}(b) and (c), which further illustrate that the above characteristic phonon modes strongly modulate the electronic distribution around the covalent C-C bonds. From this perspective, the ultrasmall (3,0) CNT can be viewed as a superconducting unit, in which characteristic collective phonon modes generate strong EPC through large phonon-induced modulations of the electronic states.

Interestingly, similar physics has also been discussed in several representative phonon-mediated superconductors. For example, strong EPC in MgB$_2$ has been associated with the stretching vibrations of B-B hexagons~\cite{PhysRevB.65.132518}, while the breathing and tilting vibrations of hydrogen cages in H$_3$S~\cite{PhysRevB.108.094519} and the breathing vibrations of Au-H octahedra in Li$_2$AuH$_6$~\cite{PhysRevB.111.L140501} also exhibit exceptionally strong EPC. Although the dominant electronic states involved in EPC differ substantially among these systems, they all host characteristic collective phonon modes capable of generating strong lattice-induced modulations of the electronic states. These results suggest that searching for structural motifs supporting such characteristic phonon modes with strong EPC provides an effective route for discovering new phonon-mediated superconductors.

\textit{Superconductivity of ultrasmall (3,0) carbon nanotube}. The phonon modes with strong EPC identified above naturally suggest robust phonon-mediated superconductivity in the ultrasmall (3,0) CNT. Using the McMillan-Allen-Dynes formula~\cite{PhysRevB.6.2577,PhysRevB.12.905}:
\begin{align}
T_c = \frac{\omega_{log}}{1.2}{\rm{exp}}[\frac{-1.04(1+\lambda)}{\lambda(1-0.62\mu^*)-\mu^*}],
\end{align}
the superconducting $T_c$ of the undoped ultrasmall (3,0) CNT is estimated to be $\sim$ 18.13 K when Coulomb pseudopotential $\mu^*$ is set as 0.10.

\begin{figure}[t]
\centering
\includegraphics[width=8.6cm]{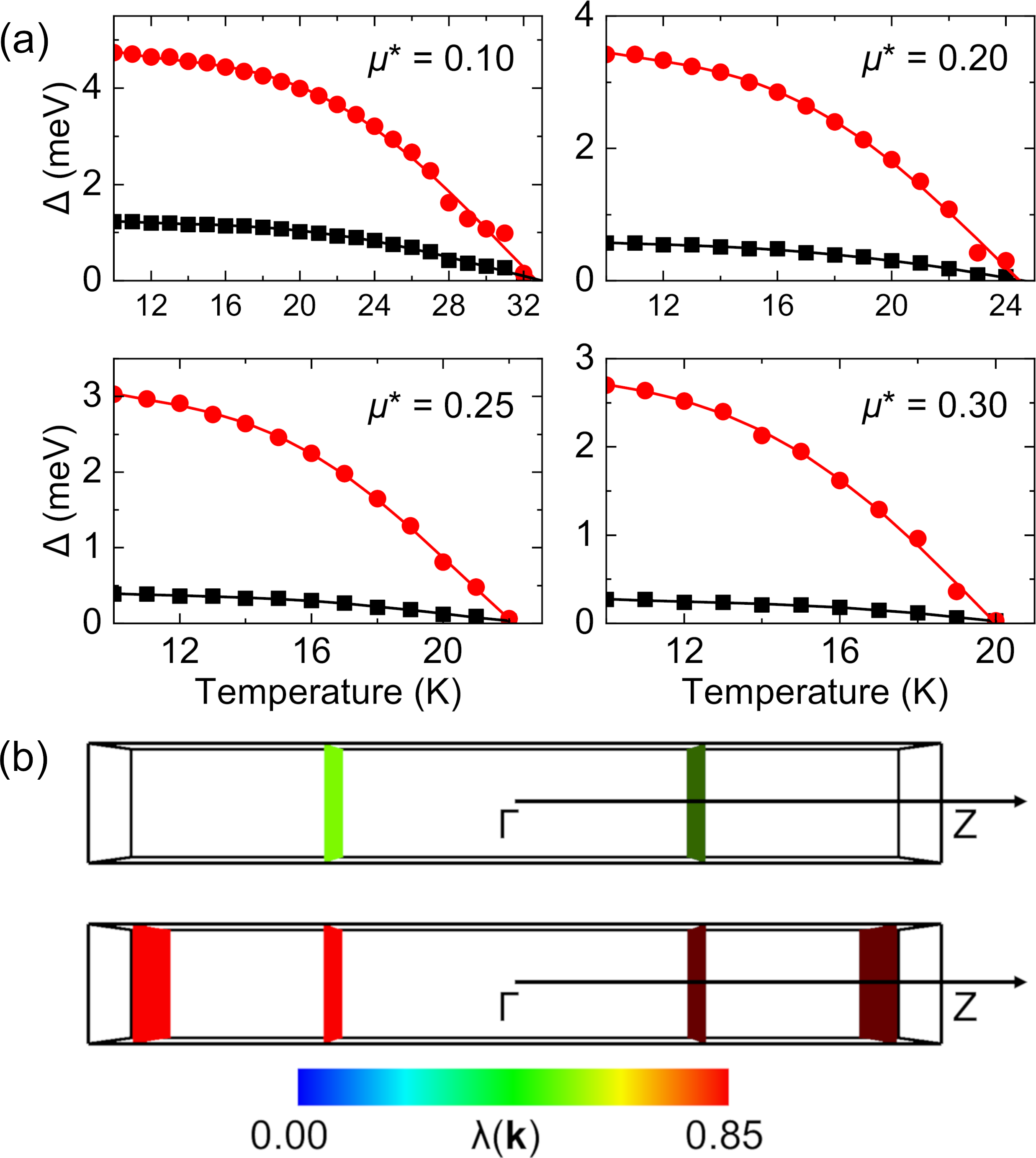}
\caption{(a) Superconducting gap $\Delta$ of undoped (3,0) nanotube. The solid lines are fitting results. (b) Fermi points with EPC $\lambda$($\bf{k}$) of undoped (3,0) nanotube. The Fermi points in the upper and lower panels originate from two different bands.}
\label{fig:Fig5}
\end{figure}

Since the McMillan-Allen-Dynes formula neglects the momentum dependence of the EPC interactions, we further solve the anisotropic Migdal-Eliashberg equations to investigate the superconducting properties of the ultrasmall (3,0) CNT. In addition, the reduced electronic screening in low-dimensional systems may lead to relatively large Coulomb pseudopotentials $\mu^*$. As shown in Fig.~\ref{fig:Fig5}(a), the calculated superconducting transition temperatures are approximately 33, 24, 22, and 20 K for $\mu^*$ = 0.10, 0.20, 0.25, and 0.30, respectively. Even for relatively large $\mu^*$, the ultrasmall (3,0) CNT still exhibits robustly high superconducting $T_c$ driven by the characteristic phonon modes with strong EPC.

As shown in Fig.~\ref{fig:Fig5}(b), the momentum-resolved EPC further reveals distinct coupling strengths on different Fermi points, exhibiting a two-gap superconducting feature in the undoped ultrasmall (3,0) CNT. Compared with previous theoretical studies on pristine one-dimensional CNT superconductors, the calculated $T_c$ of the pristine ultrasmall (3,0) CNT is among the largest reported values under ambient-pressure conditions. These results further support the picture that ultrasmall CNTs can serve as superconducting units in which characteristic phonon modes with strong EPC drive robust phonon-mediated superconductivity.

\textit{Discussion and summary.} Some theoretical studies have suggested that superconductivity in CNTs can be enhanced through various external modulations, including chemical doping~\cite{doi:10.1021/acs.jpclett.1c02000,cpl_43_1_010706}, hydrogenation~\cite{cpl_43_1_010706}, and strain engineering~\cite{li2026firstprinciplespredictionhightemperaturesuperconductivity}. However, since ultrasmall CNTs themselves are already difficult to synthesize and stabilize experimentally, further introducing external modulations may be more challenging. Here, our studies suggest that the pristine one-dimensional (3,0) CNT, serving as a superconducting unit, exhibits robust phonon-mediated superconductivity under ambient pressure without requiring additional tuning. Moreover, a theoretical study~\cite{PhysRevB.111.134518} on elemental boron shows that a new superconducting boron allotrope is constructed by taking B$_4$ tetrahedra as building blocks. From this perspective, it is natural to expect that using ultrasmall (3,0) CNTs as building blocks may achieve similar results.

Experimentally, realizing an isolated freestanding (3,0) CNT remains challenging because of its extremely small diameter, which makes substrate/host interactions, aggregation, and chirality-selective purification particularly important. However, nanoscale confinement provides a practical alternative to completely isolated tubes. In particular, zeolite-templated growth~\cite{Tang2001} has enabled the stabilization of ultrasmall CNTs in sub-nanometer channels. After properly accounting for possible effects, such as intertube coupling and doping, the confined or array-based configurations could provide a feasible platform for testing the superconductivity of freestanding (3,0) CNT predicted here.

In addition, we also notice that a recent experimental study~\cite{wang2026signatureshightemperaturecollectiveelectronic} reported possible high-$T_c$ (exceeding 200 K) superconductivity in boron-doped three-dimensional networks constructed from ultrathin (3,0) and (2,1) CNTs under ambient pressure. Our results here suggest that the superconducting $T_c$ of the pristine one-dimensional (3,0) CNT is $\sim$ 33 K ($\mu^*$ = 0.10) under ambient pressure. This difference reveals an extensive scope for future research. If more experimental reports were later confirmed, some additional mechanisms, such as intertube couplings, may be involved, which may give rise to a higher superconducting $T_c$ in three-dimensional CNT networks. Nevertheless, it should be noted that we focus on the (3,0) CNT as a representative ultrasmall CNT to highlight the concept of ``superconducting unit'' in this work, which would motivate further investigations of other ultrasmall CNTs and improve the broader understanding of superconductivity in these low-dimensional carbon systems.

In summary, we study the electronic structures, lattice dynamics, and EPC properties of the ultrasmall one-dimensional (3,0) CNT. Our calculations identify characteristic collective phonon modes with strong EPC in the ultrasmall (3,0) CNT, leading to robust phonon-mediated superconductivity under ambient pressure. Importantly, our results show that although hole doping enhances the DOS near the Fermi level, it simultaneously hardens the characteristic phonon modes with strong EPC, and ultimately reduces the total EPC constant $\lambda$ of the system. These findings highlight the central role of phonons in determining superconducting properties in ultrasmall CNTs. From this perspective, the ultrasmall (3,0) CNT can serve as a superconducting unit, namely, a structural motif hosting the phonon vibrations that can strongly modulate low-energy electronic states and drive strong EPC. Our work provides a possible strategy for discovering new phonon-mediated superconductors by identifying superconducting units that host phonons with strong EPC.

\begin{acknowledgments}
This work was supported by the National Key R\&D Program of China (Grant No. 2024YFA1408601) and the National Natural Science Foundation of China (Grant No. 12434009). 
M.G. was supported by the Ningbo Young Scientific and Technological Innovation Leaders Program (Grant No. 2023QL016).
K.L. was supported by the National Key R\&D Program of China (Grant No. 2022YFA1403103) and the National Natural Science Foundation of China (Grant No. 12174443).
Z.Y.L. was also supported by the Innovation Program for Quantum Science and Technology (Grant No. 2021ZD0302402). 
Computational resources were provided by the Physical Laboratory of High Performance Computing in Renmin University of China.
\end{acknowledgments}

\bibliography {CNT}

\end{document}